\documentclass[conference]{IEEEtran}
\IEEEoverridecommandlockouts

\ifCLASSOPTIONcompsoc
  \usepackage[nocompress]{cite}
\else
  \usepackage{cite}
\fi

\usepackage[utf8]{inputenc}
\usepackage{chngcntr}
\usepackage[cmex10]{amsmath}

\usepackage[T1]{fontenc}
\usepackage{lettrine}
\usepackage{amsfonts}
\usepackage{amssymb}
\usepackage{graphicx}
\usepackage{csquotes}
\usepackage{mathtools}
\usepackage{bbm}
\usepackage{physics}
\usepackage[normalem]{ulem}
\usepackage{tabularx}
\usepackage[nottoc]{tocbibind}
\usepackage{float}
\usepackage[caption = false]{subfig}
\usepackage{enumerate}
\usepackage[dvipsnames]{xcolor}
\usepackage{tikz}
\usepackage{pgfplots}
\tikzset{>=latex}
\usepackage{algorithm}
\usepackage[noend]{algpseudocode}
\algnewcommand\algorithmicforeach{\textbf{for each}}
\algdef{S}[FOR]{ForEach}[1]{\algorithmicforeach\ #1\ \algorithmicdo}

\usepackage{microtype}
\usepackage[shortcuts,acronym]{glossaries}
\usepackage{hyperref}
\usepackage[nameinlink]{cleveref}

\usetikzlibrary{shapes, arrows, positioning, external}
\pgfplotsset{compat=1.15}
\usepgfplotslibrary{groupplots}

\title{The Impact of Quantum Memory Quality on Entanglement Assisted Communication}
\date{}

\newacronym{bec}{BEC}{Buffered-Entanglement Channel}
\newacronym{eb}{EB}{Entanglement Buffer}
\newacronym{becn}{BECN}{Buffered Entanglement Channel Network}
\newacronym{cptp}{CPTP}{Completely Positive Trace Preserving}
\newacronym{gewi}{GEWI}{Generate Entanglement While Idle}

\author{
\IEEEauthorblockN{Stephen DiAdamo and Janis N\"otzel }\\
\IEEEauthorblockA{\textit{Emmy-Noether Gruppe Theoretisches Quantensystemdesign}\\
\textit{Technische Universit\"at M\"unchen}\\
\textit{\{stephen.diadamo, janis.noetzel\}@tum.de}}
}

\begin{document}
\maketitle

\begin{abstract}

This work explores entanglement-assisted communication, where quantum entanglement resources enable the transmission of classical information at an enhanced rate. We consider a scenario where entanglement is distributed ahead of time based on network traffic levels, and simulate a setting where idle nodes generate and store entanglement to later transmit messages at an accelerated rate. We analyze this communication model using noise models for quantum memory in various scenarios, and extend our investigation to a quantum-enhanced distributed computing environment, where entanglement storage enhances data transmission rates for cooperative data processing. We propose a protocol and demonstrate a distributed version of unsupervised clustering. Our results show that, for qubit channels, high rates of entanglement generation and modest storage requirements can surpass the classical limit with entanglement assistance.

\end{abstract}


\section{Introduction}

The field of quantum networking is becoming an increasingly important topic as quantum hardware technologies begin to perform what has been predicted theoretically. Theoretically, networked quantum technologies offer a variety of use cases that promise to outperform current and potentially any future classical implementation. Such use-cases are in distributed quantum computing~\cite{diadamo2021distributed, cuomo2020towards, caleffi2022distributed}, distributed quantum sensing~\cite{zhang2020distributed}, quantum key distribution \cite{pirandola2020advances}, and especially related to this work, quantum communication~\cite{bassoli2021quantum, notzel2020enhanced, khatri2020principles}. On the hardware side, experimentally we have seen non-local control gates between physically separated quantum processors~\cite{daiss2021quantum}, deployed quantum key distributed networks~\cite{chen2021integrated}, and recently an entanglement-assisted communication experiment~\cite{hao2021entanglement}. These results all give promise to the future goal of deploying networked quantum technologies. 

In this work, we focus on entanglement-assisted classical communication over quantum channels. At a high level, a quantum channel is a communication channel that can transport quantum systems from one physical location to another. Possible realizations of quantum channels are fiber optic cables or free space. Quantum systems can have a unique property which is not seen in purely classical states called entanglement. Two or more quantum systems being entangled means, irrespective of physical distance, the systems are correlated in their measurement outcomes. In particular, the exact form of the correlation can depend on local actions performed by the parties holding parts of the entangled systems. The effect of this behavior to network coding situations has been investigated in recent works \cite{c1,c2,c3,c4}. Entanglement-assisted communication is the communication scenario where, between communicating parties, entanglement resources are established before messages are transmitted, and at the time of message transmission, messages are encoded into sending party's part of the entangled system to then be transmitted over a quantum channel. The receiving party decodes the incoming quantum system with the assistance of their part of the entangled resource by measuring the transmitted system from the sender and their  entangled part together in a particular way. Entanglement-assisted communication has been proven in a variety of communication scenarios to increase the classical capacity of a quantum channel~\cite{bennett2002entanglement, guha2020infinite}. The effect of this additional resource on the link layer of a communication link has been also investigated in \cite{notzel2020entanglement}.

Most important to any entanglement-assisted communication scenario is the ability to distribute entanglement so that when a message is transmitted, the required entanglement resources are available at both ends of the transmission. One can consider three possible ways to achieve this: 1) Generate entanglement at the time it is needed and then use it immediately to transmit a message, 2) Generate entanglement before it is needed and store it for later use, or 3) A hybrid of the two. On one hand, generating entanglement at the time it is needed adds a layer of communication complexity to the overall protocol, but reduces the need for highly robust quantum storage devices. On the other hand, establishing the entanglement ahead of time reduces the communication complexity at the time of transmission, but storing entanglement resources requires a robust quantum memory---a highly-challenging engineering problem to build.

In this work, we, therefore, target the possibility of implementing a classical communication system with the assistance of quantum entanglement. We focus on the physical layer of the system, and specifically on the effects noisy quantum memories have on the transmission and error rates of the communication. We start with a point-to-point setting and further analyze a network of devices. As a final part of our analysis, we consider using entanglement-assisted channels for performing distributed computing. We find that with qubit channels, an advantage can be seen with noisy memories, but with relatively long coherence times. For a scheme to move from theory to practice, the hardware parameters must be considered to determine if there are advantages to gain. In this work, we aim to add clarity to the questions regarding what quality of quantum memory is necessary to perform entanglement-assisted communication. 

\section{Review of Related Work}

Past work related to the topics covered here is that which is related to quantum channels that have a queuing aspect, co-dependent queues, entanglement distribution protocols, or trading off entanglement resources for achieving entanglement-assisted capacities. In this section, we review various results and discuss their relation to the present work. Overall, this article offers a unique perspective to analyzing how building-up entanglement resources, while the network traffic rates are low, provides a communication advantage in various network settings.

We begin by reviewing the findings of Mandayam et al. \cite{mandayam2020classical}. Their study focuses on a communication scenario involving a sequentially processed stream of qubits. Before processing, these qubits are stored in a quantum memory, where processing entails performing quantum operations on the qubits or transmitting them over a quantum channel. During the period between when qubits enter the memory and when they are processed, they experience decoherence, resulting in the loss of information over time. The authors model the system after a single server queue and derive an expression for the classical capacity in terms of the qubit's waiting time. Their analysis is conducted on a queuing model that allows for unlimited storage space, with arriving qubits processed in a First-Come-First-Served order. The waiting qubits' decoherence is modeled as a quantum channel, which depends on their waiting time.

The main outcome of their work is a formula for the classical capacity of the quantum-queue channel. However, there are a few distinctions between their findings and our present study. Firstly, the queued quantum systems we consider are exclusively used for storing entanglement. For actual data, we employ a classical message buffer that stores classical bits of information, which are subsequently encoded in an entanglement-assisted manner using the stored entangled quantum system. The encoded quantum system is then transmitted over a quantum channel and decoded by the receiver. In our approach, we utilize the entanglement-assisted classical capacity of the channel to achieve accelerated communication rates. This type of capacity differs from the stricter classical capacity discussed in the previous study.

A network of nodes that can build up and re-distribute entanglement among themselves is analyzed in \cite{notzel2020enhanced}. In this work, various network topologies are considered, and a method of entanglement generation is proposed to maintain synchronized quantum buffers. This work deals with the specific case of noiseless qubit storage and superdense coding between the network nodes. Moreover, the work focuses mainly on the network layer of the stack. In this work, we take a similar approach, building on the results of \cite{notzel2020entanglement}, where we use the protocol proposed but and focusing on the physical layer aspects of the system in this case. In \cite{notzel2020entanglement}, we proposed a link-layer approach for storing and using entanglement depending on the required rate of transmission. In that work, we did not consider in depth how the physical layer effects the communication rate. Here we focus on those aspects and in particular on how memory noise of the quantum storage affects the communication rate. 

A method for using quantum channels to transmit a mixture of classical information and entanglement is analyzed in \cite{wilde2012quantum}. Here the method of \enquote{trade-off} coding is introduced which, for a given entanglement-consumption rate, some of the transmission uses entanglement-assisted coding and the remaining part of the transmission without entanglement. This method of transmitting with and without entanglement in one transmission outperforms a more common approach known as time-sharing, where some transmissions are only without entanglement assistance, and some are with it. \cite{wilde2012quantum} takes the perspective of finding the ultimate rates, but consideration for what hardware specifications are needed to execute such a protocol is not considered. In this work, our queuing model takes a time-sharing approach as a first step as our focus here is to benchmark quantum hardware for their ability to perform entanglement-assisted communication. We aim to clarify how good quantum hardware should be to perform entanglement-assisted communication before moving to the more complex coding scenario of trade-off coding.

For a communication channel using an auxiliary channel for entanglement generation along with another quantum channel for data transmission, Djordjevic has considered the transmission rates with a bosonic model \cite{djordjevic2021entanglement}. What is shown is that for such a channel, using the auxiliary channel, not for entanglement generation, but rather message transmission, the communication rates will always be higher compared to entanglement-assisted communication rates. Indeed, this work also does not consider that the entanglement can be built up and stored ahead of time. Storing quantum states for very long periods is beyond the ability of current technologies, and not making such considerations shows us that without such technology, entanglement-assisted communication may never provide an advantage over traditional methods. What we determine in this work is how much of a communication advantage is there when such memories exist, and if it is enough of a reason that additional efforts for building robust quantum memories for communication will lead to a large enough payoff. Prior work \cite{holevo2003entanglement} has pointed out the potentially infinite-fold benefits as well as the current possibilities of harvesting them \cite{guha2020infinite, sekavvcnik2022scaling}, which points to to optimistic outlook. 

The effects on entanglement concerning memory coherence times have been investigated in a simulation setting in \cite{semenenko2021entanglement}. In this work, the authors simulate entanglement sharing between two end nodes for entanglement distribution and consider an entanglement management scheme for optimizing performance. In the work, the application tested is the generation of entanglement over a linear chain of network nodes using an entanglement swapping procedure. The key metric considered is the entanglement generation rate over a system with length-dependent loss in the channel. The present work differs in that we consider a specific application requiring the consumption of entanglement for classical message transmission. In this initial work, we concern ourselves only with how the effects of memory influence the communication of classical messages.

\section{The Communication Model}

\begin{figure}
    \centering
    \includegraphics[]{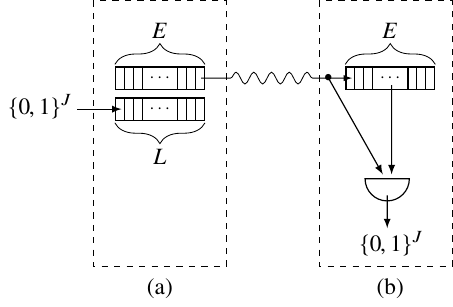}
    \caption{A communicating party in (a) has a classical message buffer of length $L$ which binary messages of length $J$ arrive and are stored. Further, at (a) is a quantum memory with $E$ slots used for entanglement buffering. Connecting parties (a) and (b) is a quantum channel, where classically encoded quantum messages are sent, depending on the state of the entanglement buffer. At (b) is another entanglement buffer that is synchronized with the one at (a). Depending on the state of the entanglement buffer, messages are either decoded with entanglement assistance or not.}
    \label{fig:ent-buff-chan}
\end{figure}

Our communication scenario is based on the second scenario discussed in the introduction, which involves generating entanglement by preparing and storing it for future use. Initially, we focus on a point-to-point communication model, as illustrated in Fig.~\ref{fig:ent-buff-chan}, where both classical and entanglement units are stored. In this model, there is a sender and a receiver who are connected by a quantum channel. They also share a synchronized quantum memory for storing entanglement resources. Additionally, the sender possesses a classical message buffer, which receives fixed-length bit-strings with a certain probability per time interval.

\begin{figure}
    \centering
    \includegraphics[]{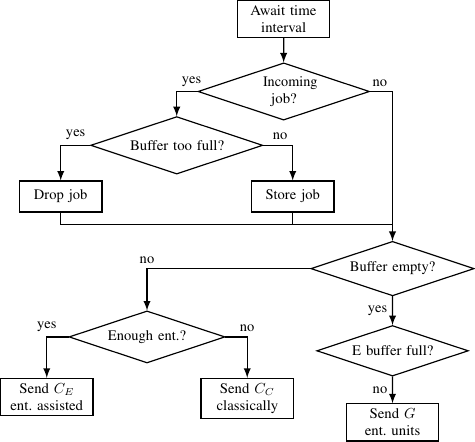}
    \caption{A flow diagram for the GEWI processes which governs an buffered entanglement channel's transmission and internal entanglement generation. The process repeats indefinitely.}
    \label{fig:bec-flow-diagram}
\end{figure}

Using this model, we can establish a communication process that prioritizes message transmission while utilizing idle periods to generate entanglement. At a high level, the process operates as follows: within a specific time interval, a classical message either arrives or does not. If a message arrives, the sender attempts to add it to a buffer of length $L$. If the buffer is too full, the message is dropped. Subsequently, the sender tries to transmit a message. In the absence of messages in the buffer, instead of remaining idle, the sender generates multiple entanglement resources, denoted as $G$, and shares half of them with the receiver, who stores them. The sender also stores the remaining half locally in an $E$-length buffer, if feasible. When both entanglement resources are available in storage and there are bits to transmit, the sender employs those resources to transmit at the entanglement-assisted capacity $C_E$ of the channel. In cases where no entanglement resources are available at the time of transmission, or if $E=0$, the sender resorts to an entanglement-free transmission, utilizing the classical capacity $C_C$ of the channel. The protocol flow is illustrated in Fig.~\ref{fig:bec-flow-diagram}, and we refer to this process, described in~\cite{notzel2020entanglement}, as the \acrfull{gewi} protocol.

\begin{figure}
    \centering
    \includegraphics[]{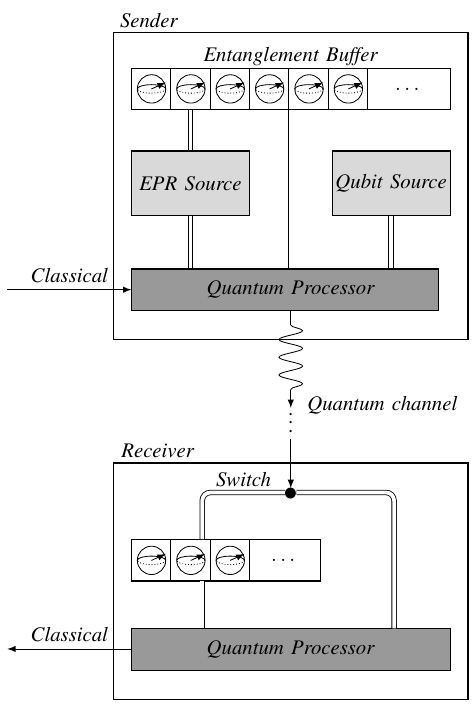}
    \caption{The sender an receiver node configuration using in the NetSquid simulations. The sender and receiver each have a quantum memory, which they maintain synchronization between. The sender has two quantum sources, one for generating qubits and one for generating EPR pairs. The sender has a classical message buffer which feeds data to the quantum processor which prepares fresh qubits from the source or qubits removed from the memory for transmission. Once transmitted, the receiver can process the qubits or store them, and outputs a classical message accordingly.}
    \label{fig:netsquid-sim-config}
\end{figure}

Based on this setup, we have devised an implementation of the system using hardware components specifically designed for quantum communication. The first requirement is the generation and storage of entanglement, which can be achieved through various methods \cite{li2020methods}, each with its own advantages and disadvantages. In our case, we assume that entanglement is generated at the sender's side, with half of it transmitted through a communication channel while the other half is stored. For storing the quantum systems, various proposals for quantum memories have been put forth in existing works \cite{simon2010quantum}. Once we have the entanglement generation and storage components in place, the remaining components involve qubit generation, establishing a channel between the two parties, and the manipulation of outgoing and incoming qubits. We depict all these components collectively in Fig.~\ref{fig:netsquid-sim-config}.

\section{Simulation Setup and Parameter Selection}\label{sec:config}

Throughout this work, we simulate the communication scenarios using the system in Fig.~\ref{fig:netsquid-sim-config} in various configurations, namely a point-to-point, unidirectional setting, a 4-node network, and a bi-directional point-to-point setting. The settings we use for simulation are a quantum channel with a classical buffer, quantum memory, and a parameter for the rate of incoming messages. For message transmission, classical information is encoded into qubits, and when possible, using pre-established entanglement, transmitted using superdense coding. When there is no entanglement available, qubits are transmitted directly with no entanglement assistance. The simulations contain a quantum channel that is modeled after a lossless $1~km$ fiber with a fixed delay time of $5~\mu s$, roughly the time light takes to travel through a $1~km$ fiber. We make the choice to use a lossless link in this case to better understand the ultimate rates for short distance communication. The qubits stored in the quantum memories experience time-dependent decoherence properties and decohere in the memories according to the memory relaxation time $T_1$ and dephasing time $T_2$. These parameters are integrated into the simulation setup directly, and the qubits stored undergo time-dependent decoherence while in storage, within the simulation setting. 

These effects act on a quantum state $\rho$ as follows, as defined in \cite[Eq. 1]{coopmans2020netsquid} and we write it here for convenience. Assuming a state is affected by the environment for  $\Delta t$ seconds, the effects are defined as follows. Firstly a quantum channel is applied to $\rho \mapsto E_0\rho E_0^\dagger + E_1\rho E_1^\dagger$, where $E_1 \coloneqq \ketbra{0} + \sqrt{1-p_1}\ketbra{1}$ and $E_1 \coloneqq \sqrt{p_1}\ketbra{0}{1}$, with $p_1\coloneqq 1 - e^{-\Delta t / T_1}$. Then, a subsequent channel $\rho\mapsto (1-p_2)\rho + p_2Z\rho Z$ is applied, where $p_2 \coloneqq \tfrac{1}{2}(1 - e^{\Delta t/T_2}e^{\Delta t / (2T_1)})$. Here we are concerned with the quality of storage required to implement the protocol and do not apply noise effects to the writing and reading steps. Loss effects for reading and writing to quantum memories will add a straightforward rate reduction proportional to the probability of loss, so we do not consider this effect here, as we are only considering storage efficiency.

The simulations we perform are concerned with how the physical layer---specifically the quantum memory---properties affect the throughput and error rate during communication. To simplify this as much as possible, we program the simulation to be independent of any link-layer protocols, but in a real communication setting it would be required for the receiver to distinguish qubits to be stored and qubits to measure using classical control information.  An initial link-layer approach for this particular communication setting has been considered in \cite{diadamo2021integrating}, where a data frame of qubits is transmitted behind a frame header, indicating how the following qubits should be treated at the receiver. Here, we ignore the link-layer protocol to focus on the ultimate rate, which can only diminished with additional link-layer processing.

To develop the simulations, we use the quantum network simulator NetSquid~\cite{coopmans2020netsquid}. NetSquid is a Python-based software library used for simulating---for one---various physical layer effects in quantum networks. Incorporated into the software are the physical models of various quantum technologies such as quantum memories, fiber optical channels, and photon sources and detectors. NetSquid is developed in a modular way, so that network nodes can contain any variation of the built-in hardware models. We construct a sender and receiver using these models. 

\section{Point to Point Channels}\label{sec:point-to-point}

In this section, we develop and present simulations of our communication model using varying quantum memory parameters and analyze the achievable throughput and decoding error rates for various scenarios. We investigate how the $T_1$ and $T_2$ times of the stored qubits and the number of memory positions available affect the throughput when using the communication setting previously defined.

To begin, each party starts with empty entanglement buffers and the sender's message buffer is empty. With some probability $r$, a bit-string of length $J$ fills a classical buffer of capacity $L$ if possible, otherwise, the message is dropped. If no message arrives, the sender uses its qubit source to generate an EPR pair, storing half of the pair in its memory and transmitting the other half over the channel. We test two approaches here, namely, if the quantum memory is full to replace the oldest qubit or to simply discard the newly generated qubit. To transmit messages, when there is no entanglement stored in the memory, 1 bit of the message is encoded into a single qubit (mapping $0$ to $|0\rangle$ and $1$ to $|1\rangle$) generated by the secondary qubit source which produces ground-state qubits and is sent over the fiber after encoding in the quantum processor. When entanglement is instead present in the memory, 1 qubit is removed from the memory using the selected queuing practice. Here we test two practices which are using a First-In-Last-Out (FILO) priority and a First-In-First-Out (FIFO) priority. The qubit removed from memory is then put into the quantum processor to be encoded and transmitted using the particular queuing practice. 

We compare the transmitted messages to those received to determine the percentage of the messages that were transmitted with and without error. To compute the average throughput, we determine the fraction of the number of bits transmitted in the total iterations multiplied by the decoding error rate. We simulate 15 different settings grouped into three different simulations. Firstly, we replicate the simulation from the previous section, using $L=J=4$ varying the $T_1,T_2$ times of the model and fixing $E=200$ memory slots. Next, we repeat the process using $J=4$ and $L=5J$. Finally, we fix the $T_1, T_2 = 1100, 1000$~ns and vary the number of memory slots in the quantum memory. The sender attempts to send a message every $10$~ns---simulating a maximum of $400$ mb/s link at most---or else generates entanglement using an entanglement source. We further assume the entanglement generation rates are equally as high.

\begin{figure}
    \centering
    \includegraphics[]{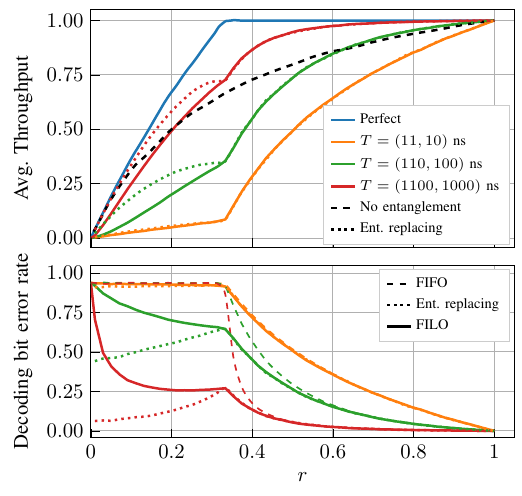}
    \caption{Two nodes separated by a fibre $1~km$ long with a fixed delay of $5000~ns$. In the first simulation, Each node has an entanglement buffer of 200 memory slots which have a specific $T_1$ and $T_2$ times associated with them. }
    \label{fig:superdense-netsquid}
\end{figure}

We analyze the results of the simulations, the first of which are plotted in Fig.~\ref{fig:superdense-netsquid}. In the upper plot is the average throughput results for varying $T_1$ and $T_2$ times of the buffer against a varying message arrival rate. As expected, the longer the $T_1$ and $T_2$ times, the smaller the error rate and thus a better throughput can be achieved. We also observe that under a certain threshold for the $T_1$ and $T_2$ times, superdense coding performs far worse than in a purely classical way, where the dashed line represents the entanglement-free case. We can see when the $T_1, T_2$ times of the memory are $11, 10$~ns respectively, since the messages are 4 bits long, until $r\approx 0.38$, the transmission produces completely random outputs with a decoding error rate of $1-1/2^4= 0.9375$. The point is exactly that at which the majority of the messages are sent using classical means over entanglement-assisted. 

In the plot of the error rates, the lower figure in Fig.~\ref{fig:superdense-netsquid}, we note that the red curve representing the $T_1, T_2$ times $1100, 1000$~ns is not non-increasing, and at around the point  $r \approx 0.38$ raises slightly to a local maximum before beginning to steadily fall. This is because the system uses FILO ordering for entanglement consumption, leaving some entangled pairs to decohere for long periods. With the red dashed line in the same plot, we show the error rates for a FIFO order. When using FIFO there is no increase in error rate as observed for FILO, but the error rates are much worse overall in comparison to FILO. The reason is that FIFO wil choose the entangled pairs that have been stored the longest when performing entanglement-assisted communication.
Especially when small amounts of data arrive, inter-arrival times for data become long and therefore the initially entangled quantum stats will typically have decohered. Another point of note is when we apply the approach of discarding old entangled pairs, replacing them with fresh pairs when the memory is full with a FILO order. We call this strategy entanglement replacing. The throughput and error rates in this case are shown with dotted lines. Here we see significant improvement at low traffic rates, eventually converging to the same trend at higher rates, as there are fewer opportunities to replace aging pairs.

\begin{figure}
    \centering
    \includegraphics{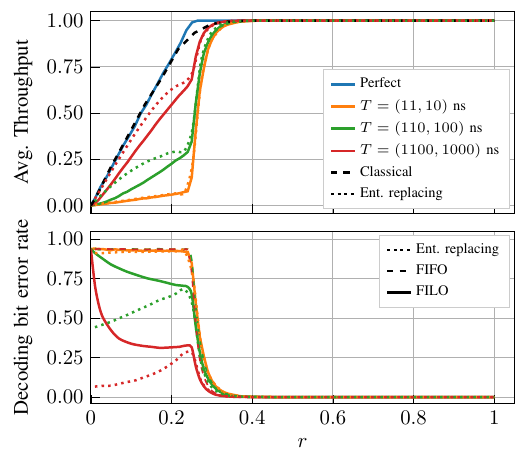}
    \caption{The same configuration as Fig.~\ref{fig:superdense-netsquid} where here the size of the message buffer here is increased to $L=5J$, allowing the message buffer to hold 5 jobs at most.}
    \label{fig:superdense-netsquid-5l}
\end{figure}

In Fig.~\ref{fig:superdense-netsquid-5l}, we repeat the simulation, but increase the message queue size so that at most 5 jobs are in the message queue at once and observe similar trends but with the plots squeezing more tightly to the smaller values of $r$. Because fewer jobs are dropped with a larger message queue size, the system is idle less often and therefore has fewer opportunities to generate entanglement. Indeed in this scenario, the average throughput in the classical case (the dashed black line) outperforms the three cases of noisy memory, which is in contrast to the single job buffer. Moreover, comparing the classical trend to the trend for the perfect quantum memory, there is only a small portion of the incoming job probability domain in which superdense coding with \acrshort{gewi} outperforms the classical case. For the error rates, we again compare the FILO and FIFO consumption of entanglement. We see again that FIFO ordering performs poorly, no better than random decoding, where FILO ordering reduces the error rates. Again, the dotted trend represents the case where entanglement bits are refreshed with a FILO ordering, and significant improvement is observed at low message arrival rates.

\begin{figure}
    \centering
    \includegraphics{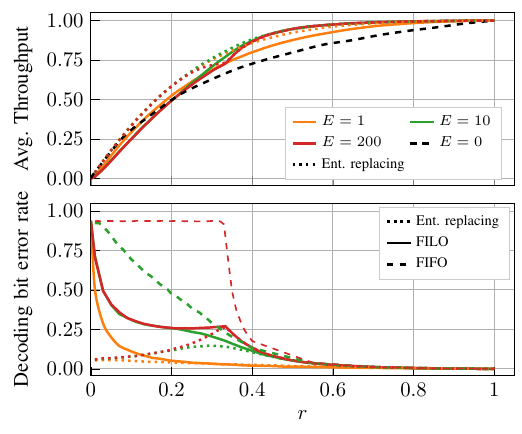}
    \caption{The same configuration as Fig.~\ref{fig:superdense-netsquid} where $T_1$ and $T_2$ times here are fixed at $1100~ns$ and $1000~ns$ respectively. The size of the entanglement buffer is varied, and the message buffer is fixed at $L = J$, allowing one job in the buffer at once.}
    \label{fig:varying-e}
\end{figure}

The final set of simulations we performed for this section is comparing the performance when varying the size of the quantum memory. In Fig.~\ref{fig:varying-e} are the trends for four memory sizes which all share the properties for $T_1, T_2=1100, 1000$~ns. What we observe is that with this choice of $T_1, T_2$ times, of all the entanglement buffer sizes that we tested, when the rate of incoming messages is low, a smaller quantum memory size performs better than a larger one, and indeed when $E=10$, until $r\approx 0.35$, outperforms the cases when $E = 200$. This is due to the aging of the entangled pairs and using a FILO consumption without replacing the older stored pairs. Since the rate of incoming messages is too low, the stored EPR pairs decohere before they can be used, whereas in the $E=10$ case, the buffer is emptied more often and so the EPR pairs are younger on average. When $r \gtrsim 0.38$, the rate at which entanglement is consumed increases enough until both $E=10$ and $E=200$ perform equivalently. With the entanglement replacing strategy, we see that for low traffic rates, replacing the entanglement with fresh units improves performance significantly, reducing error rates significantly and thus improving the overall throughput. 

Overall, from these results we can conclude the following: Firstly not only do the $T_1, T_2$ times greatly affect throughput, but also the order in which the entanglement is consumed plays a large role. By testing three orders of magnitude in this case, we can already observe $T_1, T_2$ values that perform sub-classically, and those which perform super-classically. Moreover, common orderings like FIFO and FILO are suboptimal orderings and so future work will therefore include a deeper analysis into how the ordering of entanglement consumption and the replacing of aging entanglement units optimize the throughput of the system. Second, with suboptimal entanglement consumption orderings, we see that the size of the entanglement memory plays an important role as well. With shorter $T_1, T_2$ times, using a smaller entanglement buffer will ensure the entanglement buffer is emptied more often, thereby maintaining fresher entanglement units on average and hence better decoding error rates. Lastly, in all cases, replacing stored EPR pairs with fresh pairs will improve the decoding error rates, thereby improving the total throughput.

\definecolor{green-bar}{rgb}{0.27, 0.45, 0.047}
\definecolor{blue-bar}{rgb}{0,0.117,0.647}

\section{Networks of Buffered-Entanglement Channels}\label{sec:networks}

In the previous section, we analyzed the properties of a single link under the process of \acrshort{gewi} with superdense coding, and so a next naturally arising question is: what are the properties of a network of such links? These types of questions have arisen in other queuing theory contexts, specifically, Jackson and Kelly networks for various queuing models \cite{jackson1957networks, kelly1976networks}, where the shape of the throughput can be determined analytically for some queue types. 

\begin{figure}
    \centering
    \includegraphics[scale=0.9]{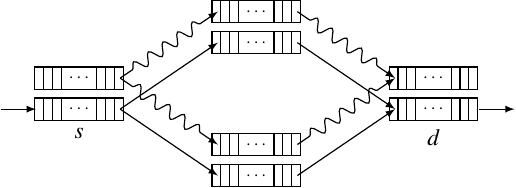}
    \caption{A network of buffered entanglement channels. Here $s$ is the source node, where classical data arrives to be sent into the network and $d$ is the sink, where classical data will be processed and removed from the network.}
    \label{fig:network-of-queues}
\end{figure}

The configuration in this simulation is the following: For the nodes with outgoing connections, transmission is performed using the process shown in Fig.~\ref{fig:bec-flow-diagram}, for each connection. When qubits are sent to a node, they are measured and re-prepared to be transmitted again depending on the state of the entanglement buffer. Each connection therefore builds up and consumes its own entanglement, with one pair of synchronized entanglement buffers per connection. In Fig.~\ref{fig:network-of-queues}, the nodes other than the source $s$ have a larger message buffer so that messages do not get dropped en route. Decoding and re-encoding at the mid-way nodes eliminates the need for entanglement swapping, simplifying the technological requirement for a physical realization in this first stage. The sink node $d$ distinguishes messages arriving from its two connections, and decodes entanglement-assisted messages based on the relay node it receives the message from. 

We use a similar settings as in the single link case: the source node can store just one message at a time with $J=L=4$ bits. We repeat the simulation with $L=5J$, allowing 5 messages in a buffer at once, but we do not show the results here due to space constraints, and that they have similar trends as the $J=L$ case. Each node has the capability to store $E=200$ entanglement units with varying $T_1, T_2$ times for the simulations. In this case, the time that the source polls for new messages is every $10~ns$, thereby simulating a maximum of $400$ mb/s network throughput.

In each of the simulations, we use a simple routing approach. We route such that a node chooses the outgoing link that has the most stored entanglement at the time of transmission. The result of this is that entanglement-assisted transmission is used as often as possible. Alternative approaches can be used, for example in \cite{notzel2020enhanced}, where the stored entanglement over entire paths is considered for routing. Given the single source of network traffic being analyzed in this case, the alternative approach is equivalent. We again bypass any link-layer protocols of the network to focus on the physical layer.

\begin{figure}
    \centering    
    \includegraphics{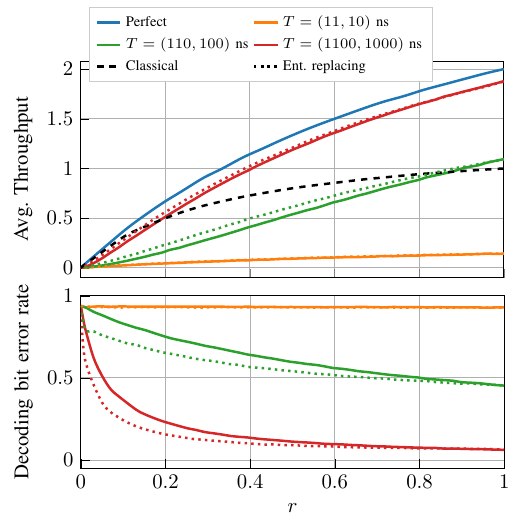}
    \caption{The throughput and error rates of a network as in Fig.~\ref{fig:network-of-queues}.}
    \label{fig:netsquid-network}
\end{figure}

For the setting when $J=L=4$, the average throughput and decoding error rate are displayed in Fig.~\ref{fig:netsquid-network}. Compared to the single-link case, the average throughput and the decoding bit error rates differ significantly compared to those in Fig.~\ref{fig:superdense-netsquid}. The main difference is, with two outgoing links from the sender, there is always one link idle during any transmission---one link is transmitting data, the other is idle. This implies that for each transmission of data, one EPR pair is generated across the idle link and therefore entanglement is always available across either the two links. Using superdense coding is therefore always possible for the first hop and an average throughput from the sender is maximized at two bits per unit time under perfect memory settings. Simulation results showed that a quantum storage of size $E = 200$ is excessive and $E = 1$ for these simulation produces the same average throughput and error plots for the different storage parameters. We therefore do not show plots for the throughput and decoding error rates for varying $E$ in this case. In the latter half of the network, the middle nodes will have the same property where one will always be idle, therefore having the ability to generate entanglement with the receiver. In this case, since $E=1$ will achieve the same throughput and average error rate, we further do not need to consider entanglement queuing practices, since for a queue with one position, all practices are equivalent. With dotted lines, we show the trend for replacing old entanglement units, which demonstrate, as with the point-to-point case in the previous section, a significant improvement in error rates and thus overall throughput.

Because entanglement is always available hop-by-hop, entanglement-assisted communication is always used, thereby leading to larger error rates with respect to the memory noise. Here we see that for $T_1, T_2=11, 10$ ns, the error rate never falls from a totally random decoding of $4$ bits, even when old entangled pairs are replaced. On the other hand, with $T_1, T_2=1100, 1000$ ns, a significant throughput advantage is observed, outperforming the classical case with relatively low incoming message rate of $r\approx0.2$. For the $T_1, T_2=110, 100$ ns case, the non-assisted data transmission rate is only surpassed at a high incoming rate of $r\approx0.9$, where entanglement will be stored for only short periods. 

What we can say in this case is that with only two connections, the use of the \acrshort{gewi} communication strategy can supersede a purely classical transmission approach even with low storage times of $T_1, T_2=110, 100$ ns for high rates of traffic, and with an order of magnitude larger, $T_1, T_2=1100, 1000$ ns, we can observe higher than non-assisted rates even with low incoming message rates. Here the size of the buffer is less important as with two links and with one source and the method of routing selected, there is more idle time on average between the two links in comparison to the single link case. In this case, we selected a network topology motivated by the results in \cite{notzel2020enhanced}, where the so-called \enquote{wide network} had the best advantage. Future work will be to consider more complex topologies. 

\section{Two-Way Communication for Distributed Computing}\label{sec:computing}

For the single-link setting, an important and natural extension to consider is a two-way communication scenario, where there are idle periods between responses between the communicating parties. Such scenarios where two parties communicate between themselves with periods of idle are common in, for example, distributed computing. In this setting, two parties send messages back and forth between themselves and once messages are received, there is a period of processing until a response message is sent. This downtime is an opportunity that can be used to build up entanglement. 

Here, we focus on (classical) distributed computation and simulate the setting where two parties perform the unsupervised learning algorithm $k$-means clustering \cite{lloyd1982least} in a distributed fashion. The initial condition of the distributed computation is that each party has access to the complete data set as well as knowledge of the initial centroid locations. For the first iteration of the algorithm, each party computes the distance to the centroids for a fair share of the data (i.e. each node has an equal number of distances to calculate), thereby labeling just their part of the data. The computed labels are then sent to the other parties, and each party receives the labels from the others to update the centroid locations. We only consider, as a proof-of-concept, binary labeling, using superdense coding to transmit the label information at an accelerated rate with two computing nodes.  The simulation setting we use in this case is much like the single link simulations in Section~\ref{sec:point-to-point}, except we duplicate the channel to accommodate two-way communication. 

The simulation setting, in this case, is the following: the nodes can communicate back and forth over a quantum channel, where all messages are transmitted via qubits. Each node can store $500$ EPR pairs in a quantum memory that has decoherence properties based on a $T_1$, $T_2$ time model, as described in earlier sections. The memory we select here is one with $T_1=1100$~ns and $T_2=1000$~ns. We repeat the process for more robust memories with $T_1=T_2=1$~ms and $T_1=T_2=10$~ms. The channel connecting the nodes, in this case, is reduced to $20$~m to simulate a data center distance, reducing the transmission latency in comparison to the previous cases. 

\begin{figure}
    \centering
    \includegraphics[scale=0.9]{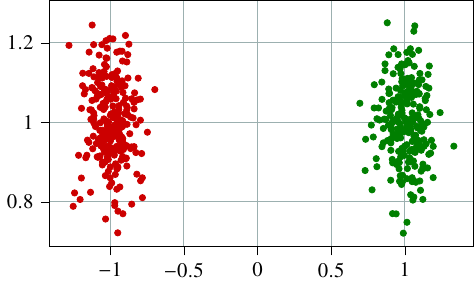}
    \caption{The synthetic data set used for the distributed clustering simulation.}
    \label{fig:scatter}
\end{figure}

The dataset we consider is a synthetic dataset consisting of 500 data points surrounding two clusters clustered around $(-1, 0)$ and $(1,0)$, shown in Fig.~\ref{fig:scatter}. The data sets are generated with a standard deviation of $0.1$ around the center. This simple configuration is chosen to focus on the communication aspect of distributed computation rather than an overly complex computation scenario. The computation and communication scenario is the following: For a maximum of 10 iterations, the two nodes compute the distances for their fair share of the 500 data points (i.e. 250 points each). During this time, they transmit a varying number of EPR pairs between themselves. We assume that the time between the last EPR transmission to the end of the data processing is 1~ms, a duration selected based on the performance of a modern laptop performing one iteration of clustering on 250 data points. During this 1~ms, the qubits decohere in the memory. After the 1~ms, the two parties share their respective 250 binary labels with the other party so that new centroid locations can be computed. In this case, this is a binary string of 250 characters. Using the preshared entanglement, they transmit several messages to the other party using superdense coding as much as possible and then revert to non-assisted message transmission when not. To mitigate errors, after each iteration of clustering, the indexes of the data points labeled by each party in the previous iteration swap, and so each party labels the data points of the other in an alternating fashion.

The metrics for performance that we consider are the total number of message transmissions used during the execution of the protocol, and the $F_1$ score of the classification comparing the final labels of the two parties, which is defined as, for a collection of binary labels $A$ and $B$, 
\begin{align}
    F_1(A, B) \coloneqq 2 \cdot \frac{precision_{AB} \cdot recall_{AB}}{precision_{AB}+recall_{AB}}.
\end{align}
We run each simulation 200 times over, plotting the average. The standard deviation in the $F_1$ score is $<2\%$ in each case, and for clarity, we do not plot error bars in the figures.

\begin{figure}
    \centering
    \includegraphics{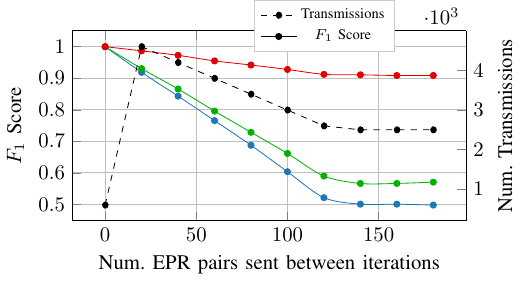}
    \caption{The trends for the total number of transmissions made between two computers over a quantum channel versus the accuracy they can achieve for the binary classification. The red (up most) plot represents $T_1= T_2 = 10$~ms, the green (middle), $T_1= T_2 = 1$~ms and the blue (lower most), $T_1, T_2 = 1100, 1000$~ns.}
    \label{fig:total_sends_vs_acc}
\end{figure}

In Fig.~\ref{fig:total_sends_vs_acc}, we see the performance of this mixed computation-communication setting. In each case, the number of transmissions follows the black dashed line, where the number of transmissions is shown on the right side of the plot. What we observe is that with a $T_1, T_2=1100, 1000$~ns (blue), the fidelity of the entanglement is not high enough at the time of transmission. As more entanglement units are used, more noise is introduced to the messages, making the decoding process nearly random, and eventually so much so, that the labels of one party and the other become completely random at the end of the 10 iterations. As one might expect, as the memory fidelity increases, the $F_1$ score improves as well, implying better matching between sender and receiver. Because superdense coding is being used to communicate, fewer and fewer transmissions are made as more EPR pairs are generated between iterations, but at the cost of a reduced $F_1$ score. When $T_1, T_2$ times is $1$~ms, still the quality of the entanglement is not sufficient, and low $F_1$ scores are observed for higher numbers of entanglement units. When the $T_1, T_2$ times is $10$~ms (red), ten times larger than the processing time, the best $F_1$ score vs. total transmissions made trade-off is seen, where there is a high $F_1$, $\geq 0.9$, score when the fewest total transmissions are used.

\section{Conclusion}\label{sec:conclusion}

In conclusion, we investigated a communication scenario where, to boost communication rates over a quantum channel, entanglement is generated during times when communication is idle to be used when communication is later required. We proposed a communication protocol that implements this and investigated the quantum memory properties required to implement the communication strategy. We considered three communication settings: a point-to-point channel, a network of four nodes with a single source and a single sink, and finally a two-party distributed computation setting. To analyze the three settings, we implemented simulations that simulate physical models of the quantum memory to determine various performance metrics for imperfect memory coherence times. We further studied how extracting qubits in a particular order affects the communication rates. What our results show is that, using our \acrshort{gewi} communication protocol, it could be possible to outperform the non-assisted communication settings where no entanglement is used, for future communication systems and distributed computing architectures. 

Indeed, these results are a starting point for this type of analysis. When considering any network protocol, it is important to also consider aspects like multi-user networks , various traffic models, different topologies, and network noise parameters can affect protocol performance. These aspects all will play a role in the overall throughput for any entanglement assisted communication network, and deeper investigation is the target for future study. 

\section*{Acknowledgments}

Funding from DFG Emmy-Noether program under grant number NO 1129/2-1 and the support of the Munich Center for Quantum Science and Technology (MCQST) are acknowledged. This research is part of the Munich Quantum Valley, which is supported by the Bavarian state government with funds from the Hightech Agenda Bayern Plus and received support by the BMBF through projects QuaPhySI, grant number 16KIS1598K, and ``Souveran. Digital. Vernetzt.'' joint project 6G-life, grant number: 16KISK002.

\bibliographystyle{IEEEtran}
% \small{\bibliography{refs}}
\small{% Generated by IEEEtran.bst, version: 1.14 (2015/08/26)

}

\end{document}